\documentclass[%
 reprint,
 amsmath,amssymb,
 aps,
showkeys
]{revtex4-1}

\usepackage{graphicx}
\usepackage{subfigure}
\usepackage{dcolumn}
\usepackage{bm}
\usepackage{afterpage}
\usepackage{amsmath}
\usepackage{amssymb}
\usepackage{mathtools}					
\usepackage{marvosym}
\usepackage{tensor}						
\usepackage{array}						
\usepackage{esvect}						
\usepackage{calrsfs}					
\usepackage{upgreek}					
\usepackage{times}						
\usepackage{helvet}						
\usepackage{courier}					
\usepackage{calligra}					

\DeclarePairedDelimiter\ket{|}{\rangle}
\DeclarePairedDelimiterX\braket[2]{\langle}{\rangle}{#1 \delimsize\vert #2}
\DeclarePairedDelimiterX\ketbra[2]{\vert}{\vert}{#1 \delimsize\rangle\langle #2}

\DeclarePairedDelimiterX\sand[3]{\langle}{\rangle}{#1 \delimsize\vert#2\vert #3}

\makeatletter
\newcommand\rcvector[2][\\]{\ensuremath{%
  \global\def\rc@delim{#1}%
    \negthinspace\begin{pmatrix}
      \rc@vector #2,\relax\noexpand\@eolst%
    \end{pmatrix}}}
\def\rc@vector #1,#2\@eolst{%
  \ifx\relax#2\relax
    #1
  \else
    #1\rc@delim
    \rc@vector #2\@eolst%
  \fi}
\makeatother
\newcommand{\colvect}{\rcvector}			

\begin{document}

\preprint{APS/123-QED}

\title{Andreev levels as a first approach to quantum computing with high Tc superconductors.}

\author{Manuel Morgado}
\author{Celso L. Ladera}%
  \altaffiliation{ Department of Physics, University Simon Bolivar, Caracas 1089, Venezuela.}
  \email[E-mail me at: ]{clladera@usb.ve}

\date{\today}

\begin{abstract}
A computing platform based on \textit{low temperature superconductors} (\textbf{LTS}) has already been proven both theoretically and experimentally. However, qubits based on high Tc superconductors (\textbf{HTS}) are not yet well understood. Here we study the \textit{Andreev bounds states} (\textbf{ABS}) in the later materials in order to show that a formal correspondence exists between the \textit{Mathieu levels} in a \textit{Cooper Pair Box} qubit built with LTS and the \textit{Andreev levels} in HTS junctions.
\end{abstract}
                           
                              
\maketitle


\section{Introduction}
Superconductivity at low temperatures has already been well explained by the BCS model, allowing us to describe phenomenae that may take place in superconducting junctions (e.g.\ the \textit{Josephson effect}) \cite{tinkham1996}. However, a satisfactory model for HTS materials has not been completely developed because the electron coupling mechanism in these unconventional superconductors is not totally understood. One of the models used in this work reaffirms that superconductivity phenomenae (e.g superconducting currents) may be explained by the \textit{Andreev reflections} mechanism \cite{kawa2010PRL}\cite{kawa2009PRB} in junctions of normal and superconducting cuprates with \textit{d-} symmetry - also known as \textit{d-wave} type superconductors. One of the experimental challenges for quantum computing with superconductors is the decoherence of LTS qubit states, and of course HTS qubits may be even more prone to high decoherence effects than LTS qubits. The high temperatures ($>77K$) in the later kind of systems are the main source of decoherence leading to very short coherence times, when compared with the expected time for a gate application to operate, the later being a key issue for us to be able to build a quantum computer based on HTS with high fault-tolerances. In  Section \ref{sec:2} we consider the symmetry of the energy gap of different kinds of superconductors, focusing our interest on the cuprates with \textit{d-wave} type symmetry and the mechanism involved in the junctions made with such materials, leading us to obtain the \textit{Andreev levels} in the spectrum of eigenenergies. Then in Section \ref{sec:3} we present a well-known theoretical model for an LTS system which lead us to a discrete nonlinear spectrum of eigenenergies that define the \textit{Mathieu levels}. In Section \ref{sec:4} we then use the two above mentioned models as a starting point to show that a formal correspondence exists between the \textit{Andreev energy levels} in HTS systems and the \textit{Mathieu energy levels} using LTS \textit{Cooper Pair Box} qubit \cite{cottet2002}\cite{koch2007PRA}. Finally, in Section \ref{sec:5}, we shall discuss how this correspondence could be interpreted and exploited, and what kind of possible difficulties would appear in actual experimental systems.
\vspace{0.3cm}
\section{HTS}\label{sec:2}

\textbf{\textit{Symmetry}}\\

From  the theory of Landau for second order phase transitions the general result emerges that the order parameter which describes such transitions should transform following one of the irreducible representations of the symmetry group at high temperatures. The symmetry group that describes the superconductor state $H$ is contained in the symmetry group of normal states $G$, ($H\subset G$) such that: $G=X\times R \times U(1) \times T$ for T$>$T$_c$, where $X$ is the symmetry gauge of the crystal lattice, $R$ is the symmetry gauge of spin-rotations and $T$ is the time reversal symmetry operator. Thus, through the decomposition of the representation of $G$ in irreducible representations, the different shapes of the order parameter can be classified into many systems of pair-condensates.\\
The point-group symmetries in cuprates superconductors are well understood \cite{tsuei2009RMP}. The pertinent coupling mechanism should also be known in order to determine the pair states in a specific crystal structure. Yet, even without this information, the order parameters, that is the energy gaps can be written as linear combinations of the functions of the basis of the irreducible representation:
\begin{equation}
\Delta(k) = \sum_{\mu=1}^{dim(\Gamma^j)}
 \alpha_{\mu} \zeta_{\mu}^j(k),
\end{equation}
where $\zeta_{\mu}^j$ is the $\mu$-th element of the basis of functions of the $j$-th irreducible representation ($\Gamma^j$), and $\alpha_\mu$ is a complex constant which is invariant under the symmetry operations of the normal state group $G$. Considering the HTS cuprates, the $CuO_2$ planes form rectangular structures of $Cu-O$ with a $C_{2\nu}$ symmetry group that results in the subtraction of the $C_4$ and the $C_{4\nu}$ reflection symmetry group (reflections with respect to $x=0$ and $y=0$). Here $C_n$ is an axis of symmetry ($c$-axis) which can be used to define rotations of $2\pi/n$, $n$ is an integer.\\

\textbf{\textit{d-wave}}\\

\begin{figure}[!h]
\includegraphics[width=0.16\textwidth]{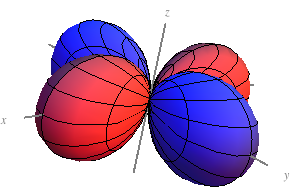}
\caption{\textit{d-wave} orbitals of the superconducting material which define the shape of the energy gap. The red and blue color show the difference of energy gap signs.}\label{orb}
\end{figure}

Recent work on the inter-laminar ac and dc Josephson effect in very high anisotropic HTS showed that it behaves as stacks of layers of $CuO_2$ coupled by Josephson interactions, which means that the superconductivity originates in the planes of $CuO_2$ \cite{kawa2011PE}. Therefore, the coupling symmetry should be reflected in the underlying symmetry of the rectangular lattice of $Cu-O$. In the schematic representation of the wave function in $k$-space there are a few candidates for the symmetry group $C_{4\nu}$. However, based on the study of Kawabata et al, we here  select  the $d-wave \,HTS$ with $d_{x^2-y^2}$ wave function shown in the figure \ref{orb}.
\\
\vspace{-0.2cm}

\textbf{\textit{Mechanism}}\\

The mechanism considered for the  $0-\pi$ transitions in Superconductor-Ferromagnetic Insulator-Superconductor ($S/FI/S$) junctions has been studied by Kawabata et al \cite{kawa2011PE} \cite{kawa2012JMMM}. There, a change of sign of the current $I_c$ may be considered as a two level system, and then in the ferromagnetic layer (ferromagnetic insulator)  excited quasi particles appear without any gap of energy leading to strong effects of decoherence.
 
In the model proposed by Kawabata for the HTS JJ, the Hamiltonian corresponding to HTS material in the junction for a 3 dimensional case is:
\begin{widetext}
\begin{equation}
\begin{split}
\mathcal{H}_{HTS} =& -t \sum_{r,r', \sigma} c^{\dagger}_{r,\sigma}c_{r',\sigma} + (2t - \mu_s) \sum_{r,\sigma} c^{\dagger}_{r,\sigma}c_{r,\sigma} +\frac{1}{2}\sum_{r} \bigg[ \Delta c^{\dagger}_{r+x,\uparrow}c^{\dagger}_{r,\downarrow} + \Delta c^{\dagger}_{r-x,\uparrow}c^{\dagger}_{r,\downarrow} +\Delta^* c_{r,\downarrow}c_{r+x,\uparrow} + \Delta^* c_{r,\downarrow}c_{r-x,\uparrow}-\\
& - \Delta c^{\dagger}_{r+y,\uparrow}c^{\dagger}_{r,\downarrow} -\Delta c^{\dagger}_{r-y,\uparrow}c^{\dagger}_{r,\downarrow} -\Delta^* c_{r,\downarrow}c_{r+y,\uparrow}-\Delta^* c_{r,\downarrow}c_{r-y,\uparrow}\bigg],
\end{split}
\end{equation}
\end{widetext}

where $\sigma$ are possible values of the spin ($\downarrow$ or $\uparrow$), $r$ is the position with coordinates $(x,y)$, $c$\,($c^{\dagger}$) are creation (annihilation) operators, $\mu_s$ is the chemical potential, $\Delta$ is the amplitude of the $d-wave$ coupling potential and $t$ is the hopping integral. For a Ferromagnetic Insulator (FI) the Hamiltonian can be written as follow:
\begin{equation}
\begin{split}
\mathcal{H}_{FI} =& -t\sum_{r,r',\sigma} c^{\dagger}_{r,\sigma}c_{r',\sigma} - \sum_r(4t-4)c^{\dagger}_{r,\uparrow}c_{r,\uparrow} +\\
&+ \sum_r (4t-\mu+V_{ex}) C^{\dagger}_{r,\downarrow}C_{r,\downarrow}.
\end{split}
\end{equation}

The external potential ($V_{ex}$) can be obtained from the tight binding model ($V_{ex}=12t+g$), $g$ is the energy gap between the states $\uparrow$ and $\downarrow$. Using the Bogoliubov transformations and the superconducting bound states ($\psi^1$ and $\psi^2$), the latter Hamiltonian could be diagonalized giving us the eigenfunctions:
\begin{equation}
\begin{split}
\psi^1_{l,m}(r) &= \Phi_1 \left[ \colvect{u, v} A e^{-iKz} + \colvect{v, u}B e^{iKz}\right]\chi_l(x) \chi_m(y),\\
\psi^2_{l,m}(r) &= \Phi_2 \left[ \colvect{v, u} C e^{iKz} + \colvect{v, u}D e^{- iKz}\right]\chi_l(x) \chi_m(y),\\
\end{split}
\end{equation}

where $A,B,C\,\text{and}\,D$ are the amplitudes of the wave functions for outgoing quasiparticles, a $\Phi_{\nu}$ is the superconducting phase:
\begin{equation}
\Phi_{\nu} = diag\left( e^{\frac{i\phi_{\nu}}{2}}, e^{\frac{-i\phi_{\nu}}{2}} \right),
\end{equation}
 
with $\nu = \{1,2\}$ and $u$,$v$ are defined as:
\begin{equation}
\begin{split}
u =& \sqrt{\frac{1}{2}\left( 1 + \frac{\Omega_{\ell m}}{E} \right)},\\
v =& \sqrt{\frac{1}{2}\left( 1 - \frac{\Omega_{\ell m}}{E} \right)}.
\end{split}
\end{equation} 

In the expressions above the energy of the channels $\ell,m$ is defined by $\Omega_{\ell m} = \sqrt{E^2 - \Delta^2_{\ell m}}$ and $\Delta_{\ell m} = \Delta (cos(q_\ell) - cos(q_m))$ where $q_\ell = \pi \ell/M+1$ and $q_m = \pi m/M+1$ ($\ell$ and $m$ are transport channels that define the corresponding energy level of the state $\psi^i_{\ell, m}(r)$), $M$ is the size of the cell studied and $E$ is the energy measured with respect to the Fermi energy. The expressions for $\chi_\ell$ and $\chi_m$ are:
\begin{equation}
\begin{split}
\chi_{\ell} (x) =& \sqrt{\frac{1}{M+1}}sin\left( \frac{\pi \ell }{M+1}\right)\hat{x},\\
\chi_m (y) =& \sqrt{\frac{1}{M+1}}sin\left( \frac{\pi m}{M+1}\right)\hat{y},
\end{split}
\end{equation}

On the other hand the expression for the \textit{wave vector} ($K$) is:
\begin{equation}
K = cos^{-1}\left( 4 - \frac{\mu_s}{2t} -C_{q_{\ell m}} - \frac{i}{2t}\sqrt{\Delta^2_{\ell m} - E^2} \right),
\end{equation}

$C_{q_{\ell m}}$ represent $C_{q_{\ell m}}= cos(q_\ell) + cos(q_m)$ and which let us to write the wave function for the FI as:
\begin{equation}\label{eq:psiFI}
\psi_{FI}(\vv{r}) = \left[ \colvect{f_1 e^{-iq_e z}, g_1 e^{-iq_h z}} + \colvect{f_2 e^{iq_e z},g_2 e^{iq_h z}} \right]\chi_{\ell}(x)\chi_m(y).
\end{equation}

where:
\small
\begin{equation}
\begin{split}
q_e =& \pi\! +\! i\left( 1 + \frac{E}{2t} + \frac{g}{4t}\!+\! C_{q_{\ell m}} \!-\! 2\cos{\!\left( \frac{\pi M}{M + 1}\right)} \right),\\
q_h =& i\left( 1 + \frac{E}{2t} + \frac{g}{4t} - C_{q_{\ell m}} - 2\cos{\!\left( \frac{\pi }{M + 1}\right)} \right).
\end{split}
\end{equation}
\normalsize

and $f_{1,2}$ and $g_{1,2}$ are the amplitudes of the wave function in the ferromagnetic insulator material.\\

\textbf{\textit{Andreev energy levels}}\\

\begin{figure}[!ht]
\centering
\includegraphics[width=0.47\textwidth]{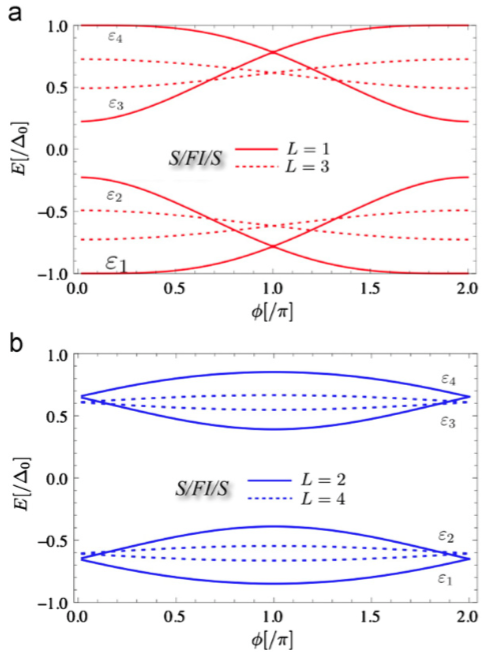}
\caption{Andreev energy levels ($\epsilon_i$): Spectra of the energies gap normalized by the gap energy in the ground state ($E/\Delta_0$) in terms of the phase function ($\phi$), for a odd number (red) and even number (blue) of Ferromagnetic Insulator material layers ($L$) obtained by Kawabata et al \cite{kawa2012JMMM}.}\label{andlvls}
\end{figure}

Using the boundary conditions for the bound states below the energy gap one obtains the \textit{Andreev energy levels} ($\varepsilon_{n,\ell,m}$). These are plotted in figures \ref{andlvls}.a and \ref{andlvls}.b. The corresponding wave functions that decay away at the interface of the junction and are given by:
\begin{equation}\label{eq:cbord}
\begin{split}
\psi_1(x,y,\lambda)& = \psi_{FI}(x,y,\lambda),\\
\psi_2(x,y,L_f+\lambda)& = \psi_{FI}(x,y,L_F+\lambda).
\end{split}
\end{equation}

In terms of these levels, the Josephson current is \cite{kawa2011PE}:
\begin{equation}
I_J(\phi) = \frac{2e}{\hbar} \sum_{n, \ell, m} \frac{\partial\varepsilon_{n,\ell,m} }{\partial \phi} f(\varepsilon_{n,\ell,m}(\phi)),
\end{equation}

where $f(\varepsilon_{n,\ell,m}(\phi))$ represents a Fermi distribution.

\vspace{0.2cm}
\section{LTS}\label{sec:3}

\textbf{\textit{Cooper Pair Box qubit}}\\

\begin{figure}[!ht]
\includegraphics[width=0.22\textwidth]{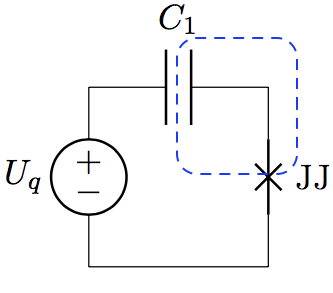}
\caption{Circuit diagram of the Cooper Pair Box. Blue dashed line points out the \textit{island}. $U_q$ represent the source, $C_1$ represent the capacitor and JJ the Josephson junction.}\label{fig:cpb}
\end{figure}

Superconducting qubits of conventional superconductors may assume several geometries and configurations that correspond to different types classes of qubits (e.g charged, fluxed and biased qubits). The well-know \textbf{CPB} qubit is composed of a Josephson junction (JJ), a capacitor (\textit{$C_1$ }) and a power source, as shown in figure \ref{fig:cpb}.\\ 

When written in terms of the charge $\hat{\mathcal{Q}}$ and phase $\hat{\mathcal{\phi}}$ operators the Hamiltonian of the CPB qubit becomes:
\begin{equation}\label{eq:cpbHam}
\mathcal{H}_{CPB} = \underbrace{\frac{\hat{\mathcal{Q}^2}}{2C_1}}_{\text{capacitor}} - \underbrace{E_{J}\cos{\left( \frac{2\pi \hat{\mathcal{\Phi}}}{\phi_0} \right)}}_{\text{JJ}},
\end{equation}
where $E_J$ is the \textit{Josephson energy}, $\phi_0$ is a reference phase. If the $\hat{n}$ is the operator of the Cooper pair number and $C_1$ the capacitance associated to the \textit{charging energy} ($E_C = e^2/2C_1$), we may write the charge operator as:
\begin{equation}
\hat{\mathcal{Q}} = 2e \hat{n} \quad \text{y} \quad \hat{\Phi} = \frac{2 \pi \hat{\phi}}{\phi_0}.
\end{equation}

Let $n_g$ be the offset number of pairs.  Replacing the CPB Hamiltonian into the eigenvalue equation $\mathcal{H}_{CPB} \ket{\psi_k(\phi)} = E_k\ket{\psi_k(\phi)}$ we obtain the equation:\\
\begin{equation}
\left[ \left( -i\frac{\partial}{\partial\phi} - n_g\right)^2 - \frac{E_J}{4E_c} \cos{\phi} \right]\psi_k(\phi) = \frac{E_k}{4E_c}\psi_k(\phi),
\end{equation}
	
which after the change of variable $g_k(x) = e^{-i(2n_g) x}\psi_k(2x)$, lead us to a \textit{Mathieu equation} \cite{abra}:
\begin{equation}\label{eq:9}
g''(x) + \left[ \frac{E_k}{E_c} + \frac{E_J}{E_c}\cos{2x} \right]g(x) = 0.
\end{equation}

Solving this equation we get the $k$-th eigenfunction $\psi_k (\phi )e^{-in_g}$ with periodical conditions:
\begin{equation}
\psi_k (\phi) = \psi_k (\phi + 2\pi)
\end{equation}
and the eigenenergies $E_k$, which in terms of the parameters $n_g$ and $E_J/E_C$ are:
\small
\begin{equation}
E_k = E_C \, \mathcal{M}_A \bigg[ k+1 - (k+1)\mod{2} + 2 n_g (-1)^{k} , -2 E_J/E_C \bigg],
\end{equation}
\normalsize
Here $\mathcal{M}_A$ is the characteristic Mathieu function and therefore:
\begin{equation}
\begin{split}
\psi_k (\phi) &= \frac{ e^{i n_g \phi} }{ \sqrt{2 \pi } } \bigg\{ \mathcal{M}_C \left[ \frac{4E_k}{E_C}, \frac{-2E_J}{E_C}, \frac{\phi}{2} \right]+\\ 
&+ i(-1)^{k+1} \mathcal{M}_S \left[ \frac{4 E_k}{E_C}, \frac{- 2 E_J}{E_C}, \frac{\phi}{2}\right] \bigg\},
\end{split}
\end{equation}

where the $\mathcal{M}_C$ and $\mathcal{M}_S$ are the \textit{Mathieu Cosine} and \textit{Sine} functions, that represent the even and odd solutions for $a$, $q$ and $z$. Simplifying the value of the parameters of the functions $\mathcal{M}_{C,S}(a,q,z)$ with $q=0$ we get:
\begin{equation}
\begin{split}
\mathcal{M}_C(a,0,z) = cos(\sqrt{a}z)\\
\mathcal{M}_S(a,0,z) = sin(\sqrt{a}z).
\end{split}
\end{equation}

Then the previous functions could be written in terms of the $\pi-0$ periodical characteristic values $mc(z)$ and $mc(z)$ of each functions in $z$:
\begin{equation}
    \begin{split}
\mathcal{M}_C(a,q,z) = mc(z)\cdot e^{irz}\\
    \mathcal{M}_S(a,q,z) = mc(z)\cdot e^{irz}
    \end{split}
\end{equation}

where $r(a,q)$ is real parameter.\\ 

\textbf{\textit{Mathieu levels}}\\

\begin{figure}[h!]
\centering
\subfigure{\hspace{-5mm}
\includegraphics[scale=0.3]{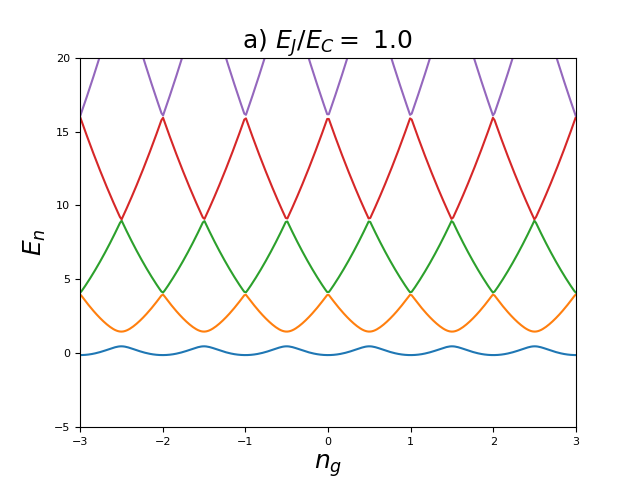}\hspace{-5mm}
\includegraphics[scale=0.3]{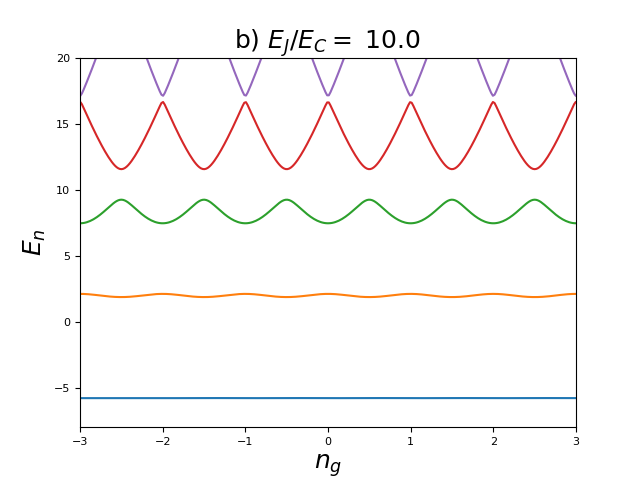}}
\subfigure{\hspace{-5mm}
\includegraphics[scale=0.3]{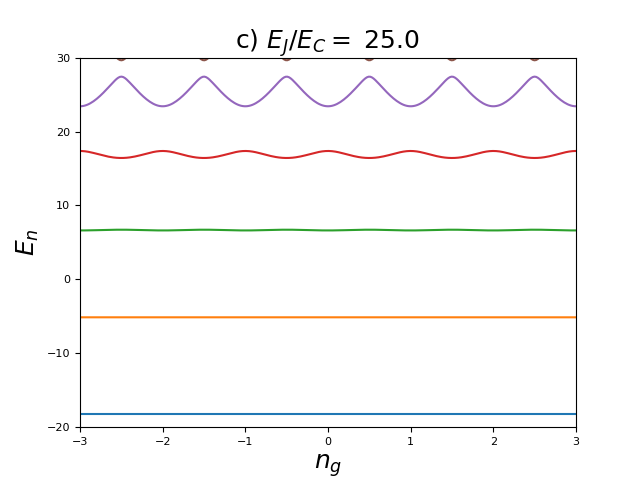}\hspace{-5mm}
\includegraphics[scale=0.3]{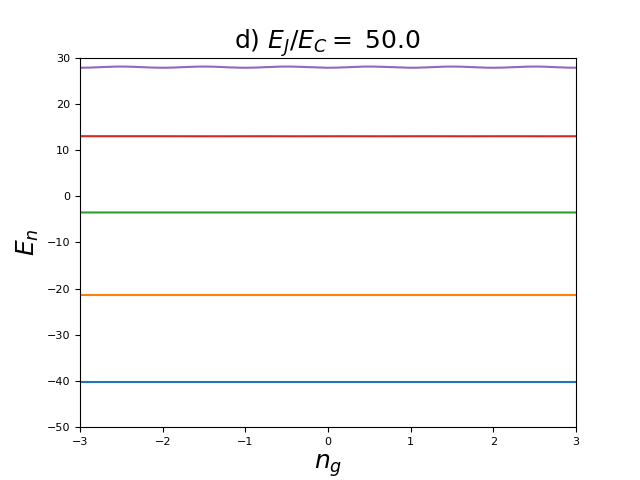}}
\caption{Non-linear spectrum of energy (first 5 levels) where is defined the \textit{Mathieu levels} in terms of $n_g$ for 4 values of $E_c/E_J= 1 , 20, 25, 50$.}\label{mathlvls}
\end{figure}

We represent the nonlinear spectra of the first five Mathieu energy levels in figure \ref{mathlvls}, using an equivalent Hamiltonian to the original (Eq.\ref{eq:cpbHam}) for different values of $E_J/E_C$ as was done by \textit{Koch et al}.\\

\begin{figure}[h!]
\subfigure{\hspace{-2mm}\includegraphics[scale=0.31]{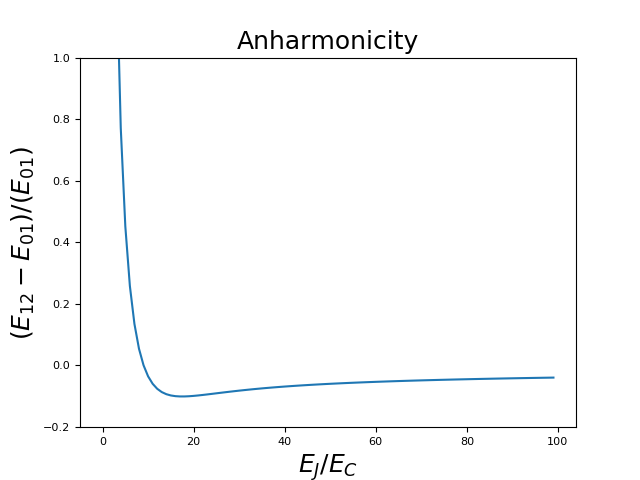}\hspace{-5mm}\includegraphics[scale=0.31]{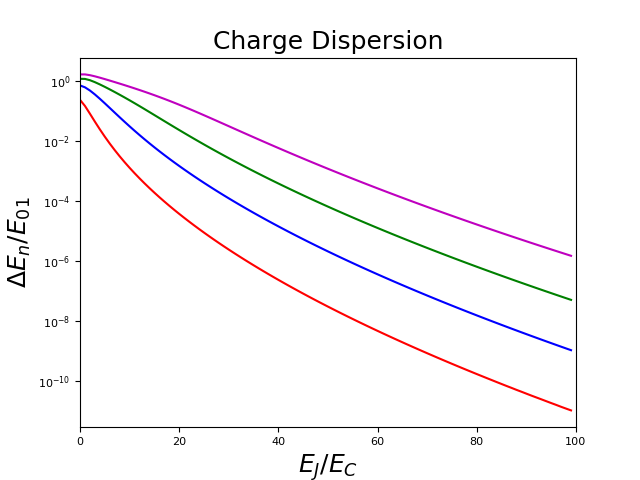}}
\caption{Anharmonicity (left), related to the coherence time (polynomial decay) and Charge dispersion (right), related to gate operation time (exponential decay). In terms of difference between the eigenenergies of the first levels.}\label{andisp}
\end{figure}

The presence of the sweet spots in these levels give us the optimal conditions that have to be set in an experiment with a CPB qubit. However, by increasing the ratio $E_J/E_C$ between the junction and charge energies we induce an  anharmonicity that does not change significantly (decays polynomially) while on the other hand the charge dispersion decrease exponentially  $\left(\propto e^{-\sqrt{8E_J/E_C}}\right)$ as it is clearly shown in the plots of figure\ref{andisp}).\\

\section{From LTS to HTS}\label{sec:4}
Let us now consider the models presented in the previous sections \ref{sec:3} and \ref{sec:4} for the case of High and Low temperatures superconductors in order to establish a correspondence between the energy levels of these two systems (HTS vs. LTS). On the one hand we have the Andreev energy levels resulting from the Andreev reflections in the JJ at the HTS sandwich; on the other hand we have the cooper pairs coming from LTS junctions that result from the BCS mechanism. This correspondence is the result of a simple comparison between the expressions of the wave function of the Ferromagnetic Insulator in the HTS system (Eq. 4 and Eq. 9) and the expression of the wavefunction in Eq. 19 for the LTS system. By the same token we may compare the boundary conditions used for the eigenfunctions used in the two cases Eq. 11 and Eq. 17. Another remarkable result is the correspondence between the eigenfunctions of Andreev model for the FI in the HTS, and the eigenfunctions of the Mathieu model in the CPB qubit using LTS (note the $\chi_\ell \, \text{and} \, \chi_m$ and the Mathieu functions $M_{C,S}$). For further details of these correspondences see below Supplemental Material \ref{supinfo}. \\

The correspondence between the levels of Mathieu at low temperatures ($E_n$) and the Andreev levels ($E/\Delta_0$) at high temperatures is graphically comparable in figures \ref{andlvls} and \ref{mathlvls}. Being both well defined two-levels systems shows that in principle it could also be possible to develop quantum devices in HTS where their qubit states could be controlled and read.

\section{Discussion}\label{sec:5}

Our comprehension of superconducting qubits constructed with conventional superconductors has made remarkable progress in the last decade, both in the experimental realm and in applications such as achieving the so-called quantum supremacy \cite{lund2017quantum}. Here, the formal correspondence between the \textit{Andreev} and \textit{Mathieu} levels has allowed us to present a correspondence between two systems made of two kinds of superconductors (HTS vs. LTS) \cite{morgado2016modelaje}. To consider these HTS materials as suitable ones for constructing devices that would represent the state of qubit will facilitate the solution of some technical problems related with the refrigeration of the quantum devices, without affecting the time of coherence ($T_1$) in relation  to the anharmonicity and to the  gate operation time, ($T_2$), the latter being related to the charge dispersion, and which is of the order of 100$\mu$s for LTS.\\

We also are aware that this approach could bring some challenges in terms of the scalability aspect as well as in terms of the decoherence due to the loss of electrons in the HTS layers. But it is important to understand that there is not general model for superconductivity that properly explains the phenomena at any temperature, which let us to hope that these challenges could be timely solved, providing us with parameters of control for such kind of systems.

\section*{Acknowledgements}

We would like to thank G. Kufatty and S. van der Woude for useful discussions.

\nocite{apsrev41Control}
\bibliography{References.bib}

\pagebreak
\widetext
\begin{center}
\textbf{\large Supplemental Material}
\end{center}
\setcounter{section}{0}
\setcounter{equation}{0}
\setcounter{figure}{0}
\setcounter{table}{0}
\setcounter{page}{7}
\makeatletter
\renewcommand{\theequation}{S\arabic{equation}}
\renewcommand{\thefigure}{S\arabic{figure}}
\renewcommand{\bibnumfmt}[1]{[S#1]}
\renewcommand{\citenumfont}[1]{#1}

\section[S.I]{Correspondence between Andreev and Mathieu energy levels}\label{supinfo}

Here we attempt to establish a correspondence between the Andreev energy levels ($A$) and the Mathieu ($M$) energy levels. It consists in defining a one-to-one correspondence between the computational basis of the two frameworks (e.g $\ket{0}_A$,$\ket{1}_A$ and $\ket{0}_M$,$\ket{1}_M$, respectively) with respect to the common computational basis ($\ket{0}$ and $\ket{1}$), this with the purpose that any unitary operation be valid in both systems, taking as a reference the LTS system. To the effect we consider Eq.\ref{eq:psiFI} that represents a superposition of the eigenstates $\ket{0}_A$,$\ket{1}_A$. These are expressed by:

\begin{equation}
\begin{split}
\ket{0}_A&= \colvect{f_1 e^{-iq_e z}, g_1 e^{-iq_h z}}\chi_{\ell}(x)\chi_m(y)=\chi_{\ell}(x)\chi_m(y)\ket{0},\\
\ket{1}_A&= \colvect{f_2 e^{iq_e z},g_2 e^{iq_h z}}\chi_{\ell}(x)\chi_m(y)=\chi_{\ell}(x)\chi_m(y)\ket{1}.
\end{split}
\end{equation}

Then we can define the following transformation $(a_{ij})_{2 \times 2}$ between the computational basis $\lbrace \ket{0}, \ket{1} \rbrace$) and the vectors that appear in the wave function of the quasiparticles in a Ferromagnetic Insulator:

\begin{equation}
\colvect{1,0} = \colvect{a_{11}\quad a_{12}, a_{21} \quad a_{22}} \colvect{f_1 e^{-iq_e z}, g_1 e^{-iq_h z}} \qquad;\qquad \colvect{0,1} =\colvect{a_{11}\quad a_{12}, a_{21} \quad a_{22}} \colvect{f_2 e^{iq_e z}, g_2 e^{iq_h z}}
\end{equation}
and therefore
\begin{equation}
\begin{split}
1-a_{11}f_1 e^{-iq_ez} &= a_{12}g_1e^{-iq_hz}\\
a_{21}f_1e^{-iq_ez} &= -a_{22}g_1e^{-iq_hz},
\end{split}
\end{equation}
and
\begin{equation}
\begin{split}
a_{11}f_2 e^{iq_ez} &= -a_{12}g_2e^{iq_hz}\\
1-a_{21}f_2e^{iq_ez} &= a_{22}g_2e^{iq_hz}.
\end{split}
\end{equation} 
And thus using fixed values for the phases of the electron and holes ($q_e \,\text{and} \,q_h$) in an specific position of the FI lattice what we get in fact is a linear transformation $[T]_{C,A}$ from the computational basis to the Andreev basis that may be written:
\begin{equation}
[T]_{C,A} = \colvect{-\frac{ g_{ 2 }e^{i(q_{h}-q_{e})z}}{g_{1}f_{2}e^{-iq_{h}z}-g_{2}f_{1}e^{i(q_{h}-2q_{e})z}} \qquad {\left( g_{ 1 }e^{ -iq_{ h }z }-g_{ 2 }\left( \frac { f_{ 1 } }{ f_{ 2 } }  \right) e^{ i(q_{ h }-2q_{ e })z } \right)}^{-1} ,-\frac { g_{ 1 }e^{ i(q_{ e }-q_{ h })z } }{ g_{ 2 }f_{ 1 }e^{ iq_{ h }z }-g_{ 1 }f_{ 2 }e^{ i(2q_{ e }-q_{ h })z } }  \qquad {\left( g_{ 2 }e^{ iq_{ h }z }-g_{ 1 }\left( \frac { f_{ 2 } }{ f_{ 1 } }  \right) e^{ i(2q_e-q_{h })z } \right)}^{-1}  }
\end{equation}

A transformation that can also be shown to be injective (one-to-one) since the spaces generated by both basis are of equal dimension. Hence, by a known theorem \cite{iri} of linear algebra it is a bijective one  or what is the same, those spaces are isomorphic. Let us now consider applying this transformation to one of the computational unitary operations say $\sigma_x$. We obtain: 
\begin{equation}
\sigma_x \ket{0}_A= \sigma_x [T]_{C,A} \colvect{f_1 e^{-iq_e z}, g_1 e^{-iq_h z}}\chi_{\ell}(x)\chi_m(y) = \chi_{\ell}(x)\chi_m(y)\sigma_x\ket{0}=\chi_{\ell}(x)\chi_m(y)\ket{1} = \ket{1}_A
\end{equation}
In the case of Mathieu levels in a LTS, where we can choose the corresponding values of the parameters $4E_k/E_C$, $E_J/E_C$ and $\phi/2$ where the energy is minimum at the first excited state, we will obtain that the $\lbrace \ket{0}_M$, $\ket{1}_M \rbrace$ present a similar relation, which must be the result of a similar analysis. Here we suggest to take the values of the parameters that correspond to the \lq\lq sweet spot\rq\rq of the CPB qubit.\\

In this way, we can say that the space of the \textit{quasiparticles} in the FI using \textbf{HTS} JJ and the \textit{cooper pairs} in the island of \textbf{LTS} JJ are isomorph, since a bijection between the \textit{Andreev energy levels} and the \textit{Mathieu energy levels} has been demonstrated to exist.

\end{document}